\documentclass[twocolumn,english,aps,manuscript,prl,reprint,longbibliography]{revtex4-1}
\usepackage[T1]{fontenc}
\usepackage[latin9]{inputenc}
\usepackage{amsmath}
\usepackage{amssymb}
\usepackage{graphicx}
\usepackage{natbib}
\usepackage{etoolbox}

\makeatletter

\usepackage[usenames,dvipsnames]{color}
\@ifundefined{textcolor}{}
{%
\definecolor{BLACK}{gray}{0}
\definecolor{WHITE}{gray}{1}
\definecolor{RED}{rgb}{1,0,0}
\definecolor{GREEN}{rgb}{0,1,0}
\definecolor{BLUE}{rgb}{0,0,1}
\definecolor{CYAN}{cmyk}{1,0,0,0}
\definecolor{MAGENTA}{cmyk}{0,1,0,0}
\definecolor{YELLOW}{cmyk}{0,0,1,0}
}
\definecolor{light-gray}{gray}{0.55}

\g@addto@macro\normalsize{%
}

\usepackage{babel}
\newcommand{\psection}[1]{\textcolor{BLUE}{\textbf{\emph{#1}}}}

\DeclareMathAlphabet{\mathpzc}{OT1}{pzc}{m}{it}
\makeatother

\usepackage{babel}
\begin{document}

\title{Spectral signatures of many-body localization with interacting photons \vspace{-5pt} }

\author{P. Roushan$^{1}$}
\thanks{These authors contributed equally to this work.}
\author{C. Neill$^{2}$}
\thanks{These authors contributed equally to this work.}
\author{J. Tangpanitanon$^{3}$}
\thanks{These authors contributed equally to this work.}
\author{V.M. Bastidas$^{3}$}
\thanks{These authors contributed equally to this work.}
\author{A. Megrant$^{1}$}
\author{R. Barends$^{1}$}
\author{Y. Chen$^{1}$}
\author{Z. Chen$^{2}$}
\author{B. Chiaro$^{2}$}
\author{A. Dunsworth$^{2}$}
\author{A. Fowler$^{1}$}
\author{B. Foxen$^{2}$}
\author{M. Giustina$^{1}$}
\author{E. Jeffrey$^{1}$}
\author{J. Kelly$^{1}$}
\author{E. Lucero$^{1}$}
\author{J. Mutus$^{1}$}
\author{M. Neeley$^{1}$}
\author{C. Quintana$^{2}$}
\author{D. Sank$^{1}$}
\author{A. Vainsencher$^{1}$}
\author{J. Wenner$^{2}$}
\author{T. White$^{1}$}
\author{H. Neven$^{1}$}
\author{D. G. Angelakis$^{3,4}$}
\author{J. Martinis$^{1,2}$}
%\email{jmartinis@google.com}

\affiliation{$^{1}$Google Inc., Santa Barbara, California, USA}
\affiliation{$^{2}$Department of Physics, University of California, Santa Barbara, California, USA}
\affiliation{$^{3}$Centre for Quantum Technologies, National University of Singapore, Singapore}
\affiliation{$^{4}$School of Electrical and Computer Engineering, Technical University of Crete, Chania, Crete, Greece }

\maketitle

\textbf{Statistical mechanics is founded on the assumption that a system can reach thermal equilibrium, regardless of the starting state. Interactions between particles facilitate thermalization, but, can interacting systems always equilibrate regardless of parameter values\,? The energy spectrum of a system can answer this question and reveal the nature of the underlying phases.  However, most experimental techniques only indirectly probe the many-body energy spectrum. Using a chain of nine superconducting qubits, we implement a novel technique for directly resolving the energy levels of interacting photons. We benchmark this method by capturing the intricate energy spectrum predicted for 2D electrons in a magnetic field, the Hofstadter butterfly. By increasing disorder, the spatial extent of energy eigenstates at the edge of the energy band shrink, suggesting the formation of a mobility edge. At strong disorder, the energy levels cease to repel one another and their statistics approaches a Poisson distribution - the hallmark of transition from the thermalized to the many-body localized phase. Our work introduces a new many-body spectroscopy technique to study quantum phases of matter.}

\psection{Introduction.} Consider a system of interacting particles isolated from the environment. Imagine it is initially prepared in a very low entropy state far from equilibrium. It is often observed that the system acts as its own thermal reservoir and approaches the equilibrium state. In this thermal phase the system shows ergodic behavior, where it uniformly explores all accessible states over time. Recent works discuss the emergence of another phase for the system in certain parameter regime where ergodicity breaks down and thermal equilibrium becomes unattainable\,\cite{Altshuler2006,Mirlin2005,Nandkishore2015,Altman2015,Yao2014,BlochMBL2015,Monroe2016,Greiner2016,GrossScience2016}. This finding is rather surprising, since intuitively one may think that interacting systems are always able to thermalize themselves. This phase is referred to as the many-body localized (MBL) phase\,\cite{Arijeet2010,Anushya2014,KnapPRL2014,Pollmann2014,Dima2015,Yasaman2015,ImbriePRL2016,Fazio2016,Antonello2017}. The conventional quantum phase transitions, e.g. from para- to ferro-magnetic, are characterized by changes in the groundstate of the system. However, the signatory differences between the thermalized and MBL phases are in dynamical behaviors, indicating that the transition involves change in the properties of all many-body eigenstates of the system. Hence the physics goes beyond the ground-state and requires study of the entire energy spectrum, which constitutes an experimental challenge.

In classical physics, the characteristic (eigen) frequencies of the system and the shape of these vibrational modes are fundamental for understanding and designing mechanical structures and electrical circuits. Similarly, in quantum physics, the quantized eigen-energies and their associated wave-functions provides extensive information for predicting the chemistry of molecules or physics of condensed matter systems. Regardless of the underlying mechanism, creating local perturbations and recording the subsequent vibrational response of the system as a function of time can reveal the characteristic modes of that system\,\cite{Roos2015}. Our method for measuring the energy spectrum of a Hamiltonian is based on this and is extremely simple. For fixed Hamiltonians, the state of a system evolves according to Schr\"{o}dinger equation
\begin{figure}
\begin{centering}
\includegraphics[width=91mm]{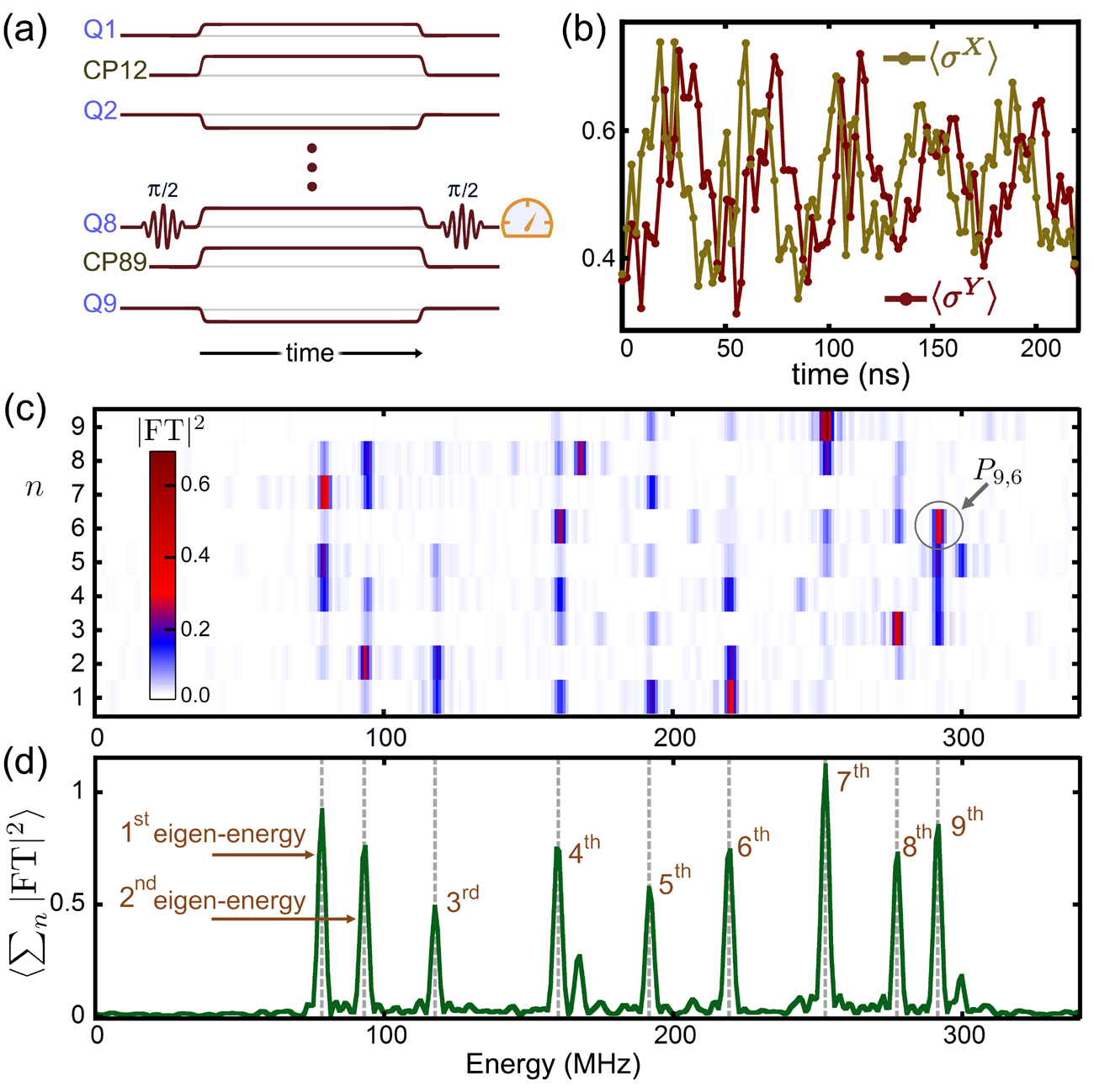}
\caption{\textbf{Time-domain spectroscopy}.\textbf{(a)} Pulse sequence used to measure eigenvalues of a time-independent Hamiltonian, Eqn.\,(2) with $J/2\pi=50$\,MHz, $U=0$, and $\mu_n/2\pi$ randomly chosen from $[0,100]$\,MHz. Initially, all the qubits are in the $|0\rangle$ state.  Using a microwave pulse, one of the qubits is then placed on the superposition of $|0\rangle$ and $|1\rangle$ state\,($Q8$ depicted here).  The coefficients in the Hamiltonian are set by applying square pulses on the qubits $\{Qn\}$ and couplers $\{CP\}$.  After the evolution, a microwave $\pi/2$ pulse is applied to the qubit in order to measure $\langle\sigma^X_n\rangle$ or $\langle\sigma^Y_n\rangle$. \textbf{(b)} Typical dataset showing $\langle\sigma^X_n\rangle$ and $\langle\sigma^Y_n\rangle$ versus time. \textbf{(c)} The FT of $\mathcal{\chi}_1(n)= \langle\sigma^X_n\rangle+i\langle\sigma^Y_n\rangle$ for $n\in\{1,2,...,9\}$. The peaks in the FT correspond to the eigenvalues of the Hamiltonian. The probability of a Fock state on $Q6$ to be in the $9^{\text{th}}$ eigenstate $P_{9,6}$ is highlighted. \textbf{(d)} Average of the FT amplitudes shown in \textbf{(c)}. Averaging is done to show all 9 peaks in one curve.}
\par\end{centering}
\end{figure}

\vspace{-17pt}
\begin{equation} \label{eq:QM}
|\psi (t)\rangle=\sum_{\alpha}{C_{\alpha} e^{-i E_{\alpha} t/\hbar}|\phi_{\alpha}\rangle},
\end{equation}

\noindent
where $E_{\alpha}$\ is an eigen-energy of the Hamiltonian and $|\phi_{\alpha}\rangle$\ is the corresponding eigenstate. Eqn.\,(\ref{eq:QM}) implies that $\{E_{\alpha}\}$ and $\{C_{\alpha}\}$ determine the frequencies and the amplitudes of the modulations in $\psi(t)$, respectively. The similarity of Eqn.\,(\ref{eq:QM}) and a Fourier transform (FT) relation suggests that the \textit{frequencies} observed in the FT of the evolution could in principle reveal $\{E_{\alpha}\}$. In addition, the \textit{magnitudes} of FT terms provide $\{C_{\alpha}\}$; these coefficients set the relative contribution of each $|\phi_{\alpha}\rangle$ to a given dynamics.

Using $9$  superconducting qubits, we constructed a 1D bosonic lattice and implement a spectroscopy method based on the above-mentioned fundamental postulate of quantum mechanics. Each of our qubits can be thought of as a nonlinear oscillator. The Hamiltonian of the chain can be described by the Bose-Hubbard model
\vspace{-8pt}
\begin{eqnarray}\label{eq:BH}
H_{BH} &=& \sum\limits_{n=1}^{9} \mu_n a^{\dagger}_{n}a_{n}+ \frac{U}{2}\sum\limits_{n=1}^{9} \,a^{\dagger}_{n}a_n(a^{\dagger}_{n}a_n-1) \nonumber\\
&+& J\sum_{n=1}^8 a^{\dagger}_{n+1}a_n+a^{\dagger}_{n}a_{n+1},
\end{eqnarray}
\noindent
where $a^{\dagger}$ ($a$) denotes the bosonic creation (annihilation) operator, $\mu_n$ is the on-site potential, $J$ is the hopping rate between nearest neighbour lattice sites, and $U$ is the on-site interaction. The qubit frequency, the nearest neighbor coupling, and nonlinearity set $\mu_n$, $J$, and $U$, respectively\,\cite{Nori2009,Nori2014,Shay2015}. In our system, we can vary the first two in ns time-scales, but $U$ is fixed.

In Fig.\,1 we show how to identify the eigen-energies of Eqn.\,(2) when it describes hopping of a single photon in a disordered potential. In the beginning of the sequence there is no photon in the system and all the qubits are in $|0\rangle$ state. Then, we place the $n^{\text{th}}$ qubit $Qn$ in the superposition of $|0\rangle$ and $|1\rangle$ state\,(Fig.\,1\textbf{(a)}). We measure the evolution of $\langle\sigma^X_n\rangle$ and $\langle\sigma^Y_n\rangle$, where $\sigma^X$ and $\sigma^Y$ are Pauli operators (acting on the $|0\rangle$ and $|1\rangle$ sub-space) (Fig.\,1\textbf{(b)}). From the $\langle\sigma^X_n\rangle$ and $\langle\sigma^Y_n\rangle$ measurements we construct $\chi_1(n) \equiv \langle\sigma^X_n\rangle+i\langle\sigma^Y_n\rangle$. Next, we vary $n$ from $1$ to $9$ to assure that the energy spectrum is fully resolved. By varing $n$ the initial states form a complete basis, and then every energy eigen-state is certain to have some overlap with one of the initial states and hence can be detected. Fig.\,1\textbf{(c)} shows the FTs of $\mathcal{\chi}_1(n)$ for each $Qn$ in which distinct peaks can be readily identified. The result of averaging the FTs is depicted in Fig.\,1\textbf{(d)}, where $9$ peaks appear and their frequencies are the $9$ eigen-energies of the Hamiltonian. The particular choices of initial states and the observables are made to avoid appearance of undesired energy peaks in the spectrum\,\cite{supp}.

\psection{Simulating 2D electrons.} Next, we demonstrate our capability to accurately set the terms in a specific Hamiltonian and resolve the corresponding eigen-energies. We simulate the problem of Bloch electrons on a 2D lattice subject to a perpendicularly applied magnetic field $B$\,\cite{Hofstadter1976,Zoller2003}. The magnetic length ($l_B=\sqrt{\hbar/eB}$) and lattice constant $a$ characterize the electron's motion, and their interplay sets the physics. The resulting energy spectrum was first calculated by Hofstadter and resembles a butterfly \cite{Hofstadter1976}. For typical crystals, the magnetic field required to 'squeeze' one flux quantum through the unit cell is of the order of several tens of thousands of Tesla, too high to be experimentally feasible. Recently, some features associated with the Hofstadter's butterfly were experimentally realized using super-lattices in graphene and cold atom systems\,\cite{Geim2013,Dean2013,Ashoori2013,Ketterle2013,Bloch2013}.

\begin{figure}[h]
\begin{centering}
\includegraphics[width=83mm]{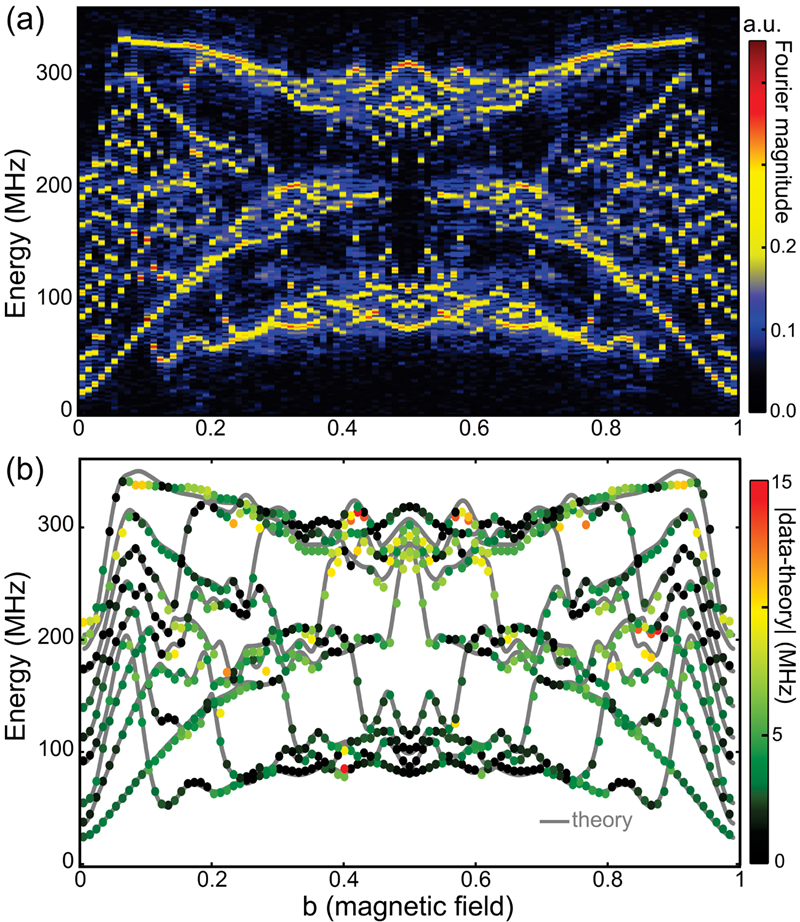} \caption{\textbf{Hofstadter butterfly}. In Eq.\,(3), we set on-site potentials $\Delta/2\pi=50$\,MHz and coupling $J/2\pi=50$\,MHz. \textbf{(a)} Data similar to Fig.\,1\textbf{(d)} is shown for $100$ values of dimensionless magnetic field $b$ ranging from $0$ to $1$. \textbf{(b)} For each $b$ value, we identify 9 peaks and plot their location as a colored dot. The numerically computed eigenvalues of Eq.\,(2) are shown with gray lines.  The color of each dot is the difference between the measured eigenvalue and the numerically computed one.}
\par\end{centering}
\end{figure}

\begin{figure*}
\begin{centering}
\includegraphics[width=182mm]{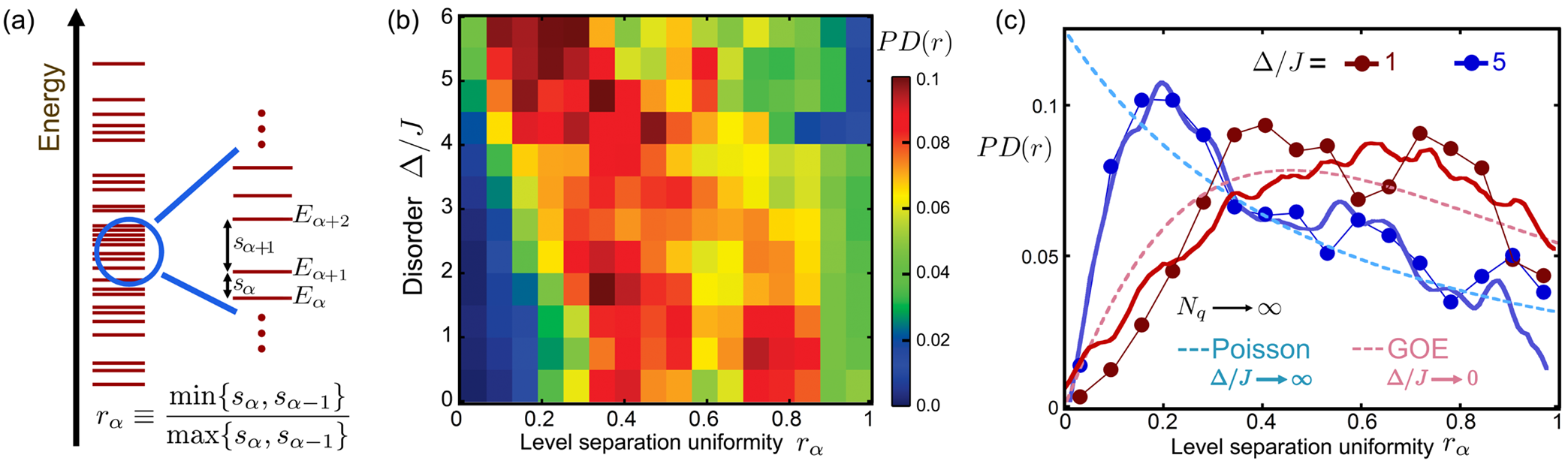}
\caption{\textbf{Level statistics and transition from GOE to Poisson}. In Eqn.\,(2), we set hopping to $J/2\pi=50$\,MHz which fixes $U/J=3.5$.  In total, 4 different irrational values of $b \in [0,1]$ are chosen and the results are averaged. \textbf{(a)} The schematic of energy levels shows how $r_\alpha$ is defined. \textbf{(b)} The measured histogram of $P(r)$ measured for various $\Delta/J$ values is presented in color. \textbf{(c)} The measured histogram $P(r)$ of $\{r_\alpha\}$ for $\Delta/J=1$ and 5. The dashed lines are plots of $P_{\text{Poisson}}$ and $P_{\text{GOE}}$ according to Eqn.\,(5), and the solid lines are numerical simulations. The change from GOE toward Poisson is indicative of vanishing of level repulsion when $\Delta$ becomes larger.}
\par\end{centering}
\end{figure*}

\vspace{3pt}
The Hofstadter energy spectra can be parameterized by a single dimensionless magnetic field $b=a^2eB/h$ which counts the number of magnetic flux quanta per unit cell. In the tight binding approximation the Schr\"{o}dinger equation takes the form of 1D Harper Hamiltonian\,\cite{Hofstadter1976}

\begin{equation}
H_{\text{Harper}}=\Delta \sum\limits_{n=1}^{9} \cos(2\pi n b)a^{\dagger}_n a_n+J \sum\limits_{n=1}^{8} a^{\dagger}_{n+1}a_n+a^{\dagger}_{n}a_{n+1}.
\end{equation}

\noindent
The $H_{\text{Harper}}$ is the special case of $H_{BH}$, reached by setting $\mu_n=\Delta \cos(2\pi n b)$ and exciting only one photon in the system, i.e. $U=0$. Note that in this limit the fermionic or bosonic nature of the particle does not matter. In Fig.\,2, we vary $b$ from $0$ to $1$ and realize $100$ different $H_{\text{Harper}}$. Similar to Fig.\,1, for each $b$ value, initial states with $n^{\text{th}}$ qubit excited are created and the evolution of $\langle\sigma^X_n\rangle$ and $\langle\sigma^Y_n\rangle$ are measured, and $n$ is varied from 1 to 9. For each $b$ value, Fig.\,2\textbf{(a)} shows the magnitude summation of the FT of $\{\mathcal{\chi}_1(n)\}$.

For large lattices with many energy levels, it is theoretically known that for rational $b$ all energy bands split into sub-bands, and for irrational $b$ the spectra become fractal and form a Cantor set. Since we have only 9 levels, what we see in Fig.\,2\textbf{(a)} are the remnants of those bands.  Nevertheless, the overall measured spectrum still resembles a butterfly. We focus on this featureful pattern of level crossings and meanderings and ask how well the measurements match simulation. In Fig.\,2\textbf{(b)}, we present the numerically computed eigen-energies with solid gray lines and the measured peaks in \textbf{(a)} with colored dots. The color of the dots shows the distance in MHz of the peaks from the simulation values. The average deviation is $3.5$\,MHz. This implies we can set the matrix elements of the Hamiltonian, which in this case includes $17$ terms, with $<2\%$ error. This unprecedented capability in controlling a large quantum system is achieved through careful modeling of the qubits as non-linear resonators.

By placing two photons in the system, we next study the simplest interacting cases\,($U \neq 0$, with no mapping to electronic system). The rest of data presented in this work is taken by using the following procedure (2-photon protocol). We realize a quasi-periodic potential by setting $\mu_n=\Delta \cos(2\pi n b)$. In total, 4 different irrational values of $b \in [0,1]$ are chosen and the corresponding results are averaged. The irrational choice of $b$ assures that the periodicity of the potential and lattice are incommensurate. In Eqn. (2), we set $J/2\pi=50$\,MHz, which results in $U/J=3.5$. The initial states are made by placing two qubits ($Qn$ and $Qm$) in the superposition of the $|0\rangle$ and $|1\rangle$ states. We measure two-point correlations and construct $\mathcal{\chi}_2(n,m)\equiv \langle\sigma^X_n\sigma^X_m\rangle -\langle\sigma^Y_n\sigma^Y_m\rangle+i\langle\sigma^X_n\sigma^Y_m\rangle+i\langle\sigma^Y_n\sigma^X_m\rangle$. The peaks observed in the FT of $\mathcal{\chi}_2(n,m)$ are the eigen-energies of $H_{BH}$ in the two-photon manifold\,\cite{supp}.

\psection{Energy level statistics in an interacting system.} Perhaps the most direct way of examining ergodic dynamics and its breakdown is by studying the distribution of the energy levels\,\cite{Huse2007,Atas2013,Bohigas1984}. Using the 2-photon protocol, we measure the evolution of $\mathcal{\chi}_2(n,m)$ for various strengths of disorder $\Delta$. We identify the peaks in the FT of $\mathcal{\chi}_2(n,m)$ as the energy levels ${E_\alpha}$. Let $s_\alpha=E_{\alpha+1}-E_\alpha$ be the nearest-neighbor spacings (illustrated Fig.\,3\textbf{(a)}), and level separation uniformity $r_\alpha \equiv \text{min}\{s_\alpha,s_{\alpha-1}\}/\text{max}\{s_\alpha,s_{\alpha-1}\}$. From our measured $\{E_\alpha\}$ we compute the associated $\{r_\alpha\}$ and construct their probability distribution ($PD$, Fig.\,3\textbf{(b)}). For low disorder, the $PD$ is mainly centered around the $r_\alpha$ values close to half, and with increase of disorder the histogram's peak shifts toward smaller $r_\alpha$ values.

It has been postulated that in the ergodic phase the statistics of levels is the same as the ensemble of real Hermitian random matrices, which follow the Gaussian Orthogonal ensemble\,(GOE)\,\cite{Bohigas1984}. In the localized phase, the energy levels become uncorrelated due to disorder and hence it is expected to show a Poisson distribution in energy landscape.  The probability distribution  of $\{r_\alpha\}$ for the ergodic and many-body localized phases, respectively, are

\begin{equation}
PD_{\text{GOE}}(r)=\frac{27}{4} \frac{r+r^2}{(1+r+r^2)^{5/2}}, PD_{\text{Poisson}}(r)=\frac{2}{(1+r)^2}.
\end{equation}

In Fig.\,3\textbf{(c)}, we focus on $\Delta/J=1$ and $5$, showing the measured histograms with dots and the numerical simulations with solid lines. The dashed lines are plots of Eqn.\,(5), providing the expected behavior in the thermodynamic limit (number of sites $N_q \rightarrow \infty$), and for limiting values of $\Delta/J$. In contrast to these cases, the finite size of our chain results in features that can be seen in both data and simulation. When disorder is small, the energy eigenstates are extended across the chain (we will show this in Fig.\,4,) and hence the energy levels repel each other. Consequently, there are strong correlations between the levels and an equidistant distribution of levels would be favorable. When $\Delta$ becomes larger, the eigenstates become localized in space and unaware of each others presence at a given energy and level repulsion ceases. Therefore, the levels independently distribute themselves, showing a Poisson distribution in the energy landscape. The exact realization of Poisson distribution takes place only when $J/\Delta \rightarrow 0$\,; in our case $J/\Delta=0.2$, which is where the peak in the histogram appears. Since the Poisson distribution is the signature of independent events, we conclude that the transition from ergodic to localized phase is associated with vanishing correlations in energy levels.

\begin{figure}
\begin{centering}
\includegraphics[width=88mm]{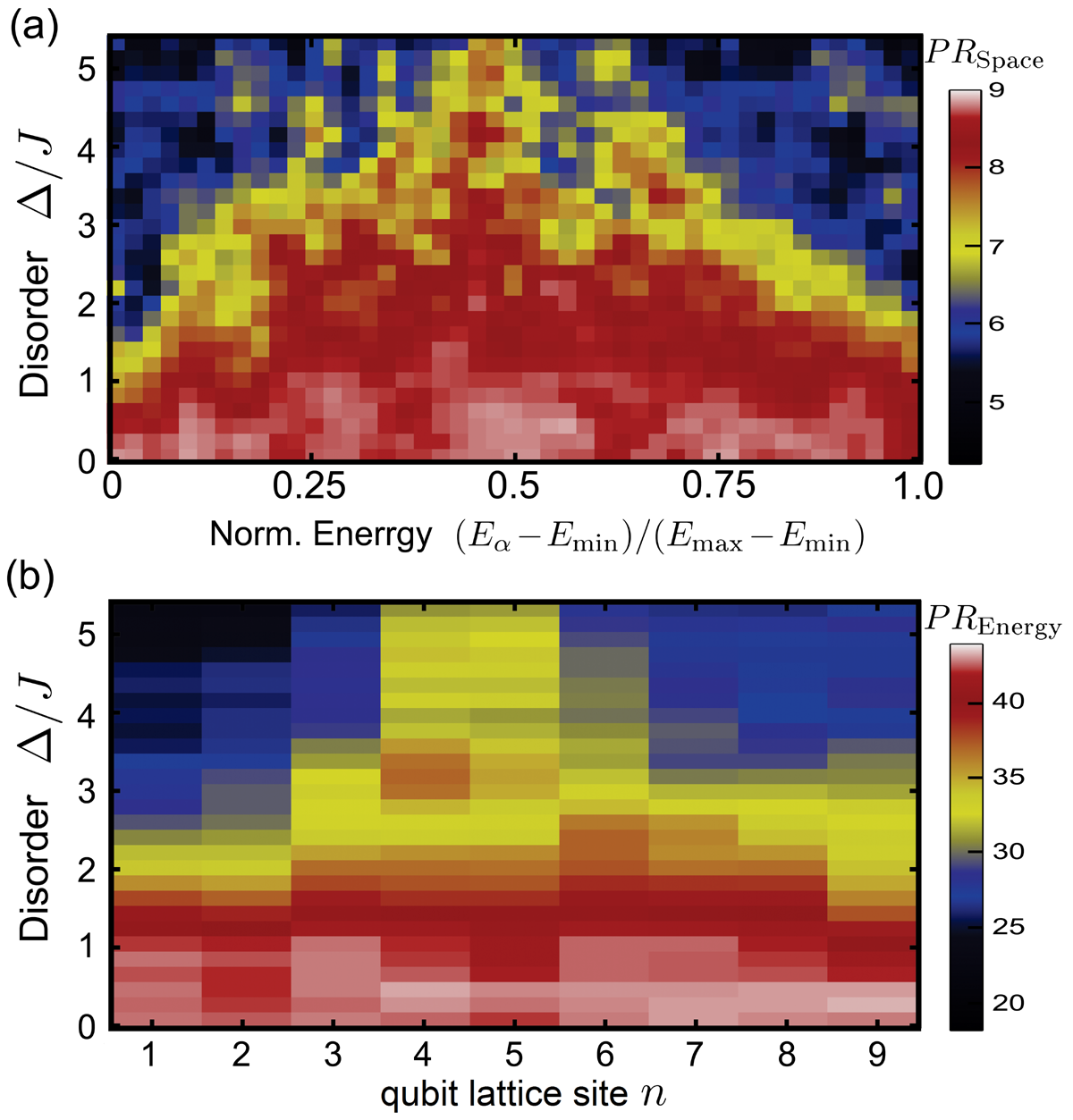} \caption{\textbf{Participation ratio and Mobility edges}. In Eqn.\,(2), we set $b=(\sqrt{5}-1)/2$, $J/2\pi=50$\,MHz, which results in $U/J=3.5$.We measure the evolution of $\mathcal{\chi}_2(n,m)=\langle\sigma^X_n\sigma^X_m\rangle +\langle\sigma^Y_n\sigma^Y_m\rangle+i\langle\sigma^X_n\sigma^Y_m\rangle+i\langle\sigma^Y_n\sigma^X_m\rangle$ for all pairs of $n,m \in \{1,2,...,9\}$ as a function time for various strengths of disorder $\Delta$. From the magnitude of the peaks seen in the FT of the data the probabilities relating the positions of two-photon Fock states to energy eigenstates $\{P_{\alpha,n}\}$ are extracted.  See\,\cite{supp} for details. The computed \textbf{(a)} $PR_{\text{Space}}$ and \textbf{(b)} $PR_{\text{Energy}}$ based on Eqn. (4) are plotted. The $E_{\text{max}}-E_{\text{min}}$ is the width of the energy band at a given $\Delta$.}
\par\end{centering}
\end{figure}

\psection{Spatial extend of eigen-energies.} A key signature of transition from ergodic to MBL phase is the change in the localization length of the system from being extended over entire system to localized over a few lattice sites. This physics can be studied by measuring the probability of each energy eigen-state being present at each lattice site $\{P_{\alpha,n}\}$\,\cite{supp}. In our method, the frequencies of the FT signal give the eigen-energies, and from the magnitude of the FT terms $\{P_{\alpha,n}\}$ can be measured. For instance, $P_{9,6}$ is highlighted in Fig.\,1\textbf{(c)}. In the study of metal-insulator transition\,\cite{The50years,Huse2007}, a common way to quantify the extension in real-space or energy landscape is via the second moment of the probabilities, defined by Participation Ratio ($PR$)

\begin{equation}\label{eq:PR}
PR_{\text{Space}}(\alpha)\equiv 1/\sum\limits_{n} P_{\alpha,n}^2,\, PR_{\text{Energy}}(n) \equiv 1/\sum\limits_{\alpha}P_{\alpha,n}^2.
\end{equation}

\vspace{9pt}
\noindent
$PR_{\text{Space}}$ indicates the number of sites over which an energy eigenstate $|\phi_{\alpha}\rangle$ has an appreciable magnitude. Similarly, $PR_{\text{Energy}}$ measures how many energy eigenstates have significant presence on lattice site $n$. Note that the first moment of the probability distributions is normalization conditions $\sum_\alpha P_{\alpha,n}=1$ and $\sum_n P_{\alpha,n}=1$.

Demonstrated that we can fully resolved the energy spectrum of the two-photon energy manifold, we now extract $\{P_{\alpha,n}\}$.  In Fig.\,4\textbf{(a)}, we compute $PR_{\text{Space}}$ for various disorder strengths and present them in the order of increasing energy. In this energy manifold, there are $36$ single \,(e.g. $|001000100\rangle$) and 9 double occupancy states\,(e.g. $|000020000\rangle$), which gives ${9 \choose 2}+{9 \choose 1}=45$ energy levels. For low disorder\,($\Delta/J <1$), $PR_{\text{Space}}$ is about 8, indicating almost all energy eigenstates are extended over the entire chain of 9 qubit lattice sites. As the strength of disorder increases, the eigenstates with their energies close to the edge of the energy band start to shrink, while eigenstates with energies in the middle of the band remain extended at larger disorders. This is consistent with the notion that localization begins at the edges of the band, and a mobility edge forms (the yellow hue) and approaches the center of the band as disorder becomes stronger\,\cite{The50years}. This is similar with the Anderson localization picture, in which localization begins at the edges of the band, and a mobility edge forms (the yellow hue) and approaches the center of the band as disorder becomes stronger\,\cite{The50years}. However, the existence of mobility edge has been theoretically questioned, and proper investigation of it requires going to larger systems and finite size scaling\,\cite{Chris2014,Alet2015,MobilityEdge2016,Antonello2017}. Given that numerical exact diagonalization is limited to small size systems, scaling up the experiment could shed light on this matter and general understanding of MBL\,\cite{Bhatt2015,Iyer2013}.

\begin{figure}
\includegraphics[width=80mm]{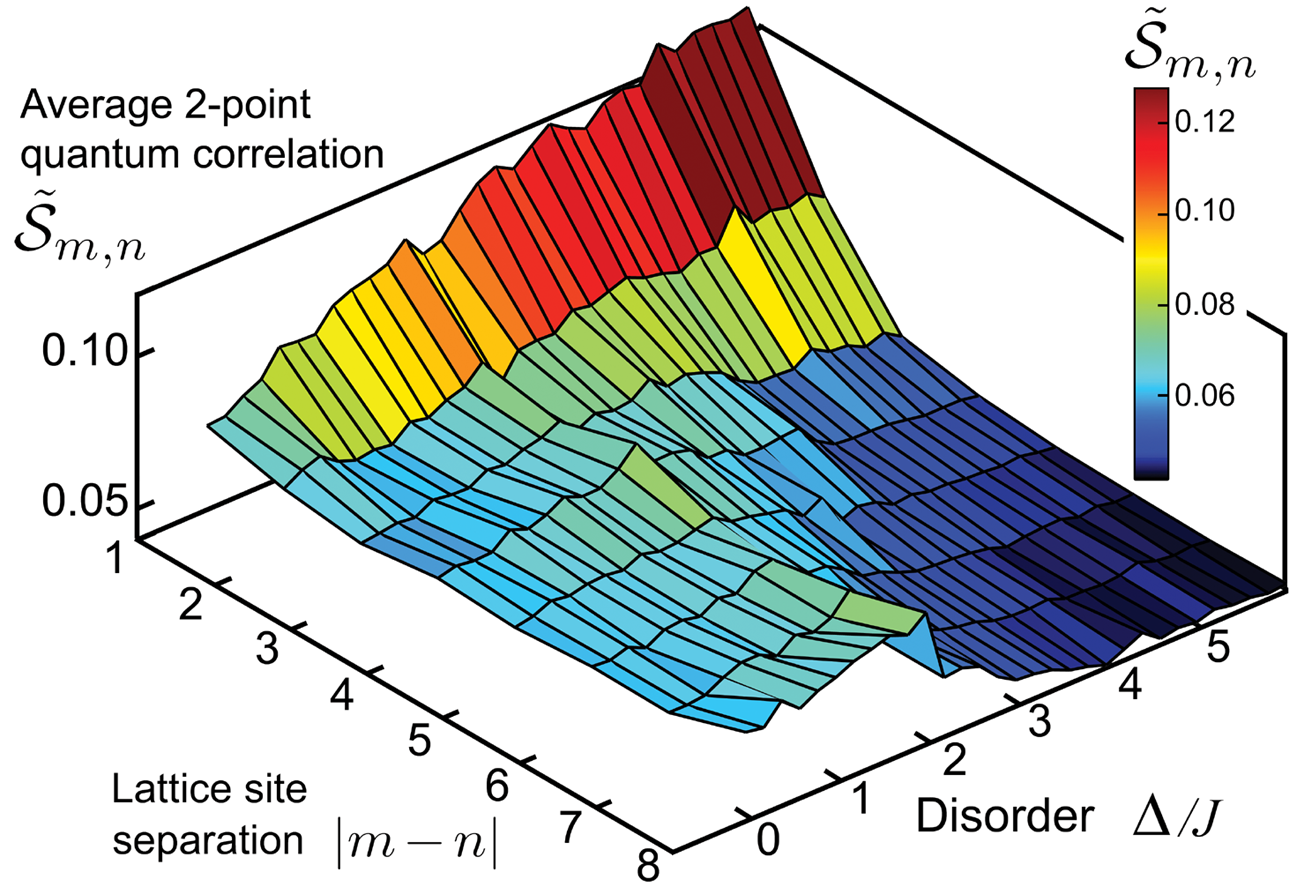}
\caption{\textbf{Quantum correlations}. In Eqn.\,(2), we set $b=(\sqrt{5}-1)/2$, $J/2\pi=50$\,MHz, and $U/2\pi=175$\,MHz. We measure $\mathcal{S}_{m,n}=|\langle\sigma_m^1 \sigma_n^2\rangle-\langle\sigma_m^1\rangle \langle\sigma_n^2\rangle|$ as a function time for various strengths of disorder $\Delta$, where $\sigma^1,\sigma^2 \in \{\sigma^X, \sigma^Y\}$ and $m,n \in \{1,2,...,9\}$. All ${9\choose 2}=36$ possible pairs of qubits are excited. The color shows $\mathcal{S}_{m,n}$ averaged over time (from 0 to 250\,ns) and combinations with the same $|m-n|$. The change of correlations from almost uniform to exponentially decaying is consistent with change in behavior from ergodic to localized.}
\end{figure}

In Fig.\,4\textbf{(b)}, we plot the $PR_{\text{Energy}}$, which shows that as the disorder becomes stronger, the number of eigenstates present at a given lattice site reduces, indicating that eigenstates are becoming localized on lattice sites. Furthermore, with increasing disorder, the eigenstates are avoiding the edges of the chain and more eigenstates have presence toward the center of the chain. The changes in $PR_{\text{Space}}$ and $PR_{\text{Energy}}$ are the fastest near $\Delta/J=2$, suggestive of a phase transition that has been smeared out due to finite size effects. Nevertheless, we emphasize that the quantum phase transition to the MBL phase is only defined in the thermodynamic limit\,($N_q \rightarrow \infty$)\,\cite{ImbriePRL2016}. Given the finite size of our system and the presence of only two interacting particles, it is interesting that we see several signatures associated with the MBL phase transition.

\psection{Quantum correlations.} To provide a comprehensive picture of the transition to the localized phase, we study two-site quantum correlations $\mathcal{S}_{m,n}$ as a function of disorder strength $\Delta$ and distance between lattice sites $|m-n|$. We measure $\mathcal{S}_{m,n}\equiv|\langle\sigma_m^1 \sigma_n^2\rangle-\langle\sigma_m^1\rangle \langle\sigma_n^2\rangle|$, where $\sigma^1,\sigma^2 \in \{\sigma^X, \sigma^Y\}$ and $m,n \in \{1,2,...,9\}$, for all $m$ and $n$ combinations  and Pauli operators. Fig.\,5\textbf{(a)} shows ${\mathcal{\tilde{S}}_{m,n}}$, computed by averaging $\mathcal{S}_{m,n}$ over time and all possible combinations with the same $|m-n|$. For $\Delta$ up to $\Delta/J \approx 2$, ${\mathcal{\tilde{S}}_{m,n}}$ is rather symmetric in $|m-n|$, and for $\Delta/J>2$ it exponentially decays with $|m-n|$. Intuitively, strong $\Delta$ creates large potential barriers that wave-functions cannot tunnel through and consequently correlations cannot develop. Interestingly, for $|m-n|<3$, as disorder becomes stronger,  ${\mathcal{\tilde{S}}_{m,n}}$ becomes larger, indicating that correlations cannot propagate far and locally build up in the potential 'puddles'. These observations are consistent with the signatures of the transitions from the metallic phase, where correlations are distance independent, to the localized phase where they decay rapidly with distance.

\psection{Conclusion.} Our work demonstrates the novel information about various phases that can be extracted if one directly resolves the energy levels of a system. Our findings signifies the generality of the MBL phenomena and the fact that its underlying physics prevails regardless of the details of the system. Our work demonstrates that novel properties of various phases can be extracted by directly measuring the energy levels of a system. It is interesting to consider the application of this method to a device with a few tens of qubits where classical simulations will begin to fail. The technique presented here is scalable to more qubits but is ultimately limited by the frequency broadening that results from decoherence. For large systems, the level spacing becomes exponentially denser and the current approach needs to be revised; promising methods are suggested in \,\cite{Nandkishore2014,Bhatt2015}. Nevertheless, the valuable computational resource that our platform offers resides in measuring the dynamics of observables and quickly becomes intractable for classical computers.

\vspace{2mm}
\footnotesize \textbf{Acknowledgments:} We acknowledge discussions with Y. Bahri, D. Huse, E. Kapit, M. Knap, L. Hormozi, O. Kyriienko, M. F. Maghrebi, V. Oganesyan, A. Scardicchio, and G. Zhu.

\vspace{2mm}
\footnotesize \textbf{Correspondence:} All correspondence should be addressed to pedramr@google.com and dimitris.angelakis@gmail.com.

\end{document}

\title{Supplementary materials:
Spectral signatures of many-body localization with interacting photons}

\author{P. Roushan$^{1}$}
\thanks{These authors contributed equally to this work.}
\author{C. Neill$^{2}$}
\thanks{These authors contributed equally to this work.}
\author{J. Tangpanitanon$^{3}$}
\thanks{These authors contributed equally to this work.}
\author{V.M. Bastidas$^{3}$}
\thanks{These authors contributed equally to this work.}
\author{A. Megrant$^{1}$}
\author{R. Barends$^{1}$}
\author{Y. Chen$^{1}$}
\author{Z. Chen$^{2}$}
\author{B. Chiaro$^{2}$}
\author{A. Dunsworth$^{2}$}
\author{A. Fowler$^{1}$}
\author{B. Foxen$^{2}$}
%\author{M. Giustina$^{1}$}
\author{E. Jeffrey$^{1}$}
\author{J. Kelly$^{1}$}
\author{E. Lucero$^{1}$}
\author{J. Mutus$^{1}$}
\author{M. Neeley$^{1}$}
\author{C. Quintana$^{2}$}
\author{D. Sank$^{1}$}
\author{A. Vainsencher$^{1}$}
\author{J. Wenner$^{2}$}
\author{T. White$^{1}$}
\author{H. Neven$^{1}$}
\author{D. G. Angelakis$^{3,4}$}
\author{J. Martinis$^{1,2}$}
\email{jmartinis@google.com}

\affiliation{$^{1}$Google Inc., Santa Barbara, CA 93117, USA}
\affiliation{$^{2}$Department of Physics, University of California, Santa Barbara, CA 93106, USA}
\affiliation{$^{3}$Centre for Quantum Technologies, National University of Singapore, 2 Science Drive 3, 117542, Singapore }
\affiliation{$^{4}$School of Electrical and Computer Engineering, Technical University of Crete, Chania, Crete, 73100 Greece }

\let\oldthebibliography=\thebibliography
\let\oldendthebibliography=\endthebibliography
\renewenvironment{thebibliography}[1]{%
    \oldthebibliography{#1}%
    \setcounter{enumiv}{5}%
}{\oldendthebibliography}

\maketitle
\textbf{}
\vspace{-20pt}
\tableofcontents
% \clearpage

\vspace{-20pt}
\section{\textcolor{SECTIONCOLOR}{1. Device: the superconducting qubits with gmon architecture}}

\begin{figure}[h]
\floatbox[{\capbeside\thisfloatsetup{capbesideposition={right,top},capbesidewidth=6cm}}]{figure}[\FBwidth]
{\caption{An optical micrograph of the device which consists of 9 qubits in a 1D chain with adjustable coupling between every pair of qubits. The qubits appear as small vertical rectangles in the middle of the chip. The couplers are the two square loops that are between the qubits. The wiring lines that are routed to the perimeter of the chip are used to control the qubits and the interaction between them. The meandering lines above the qubits are the readout resonators. The qubits are connected with an adjustable
coupler. Each qubit is a non-linear LC resonator, and the two qubits are inductively coupled through the mutual inductance to a coupler loop. The coupler loop has a single Josephson junction, which can be tuned by applying magnetic flux into the coupler loop, allowing variable coupling strength between the two
qubits in a few $ns$ time scales. For a detailed discussion of principle of operation and calibration routines, see references \cite{Yu2014,Geller2015, Ergodicity2015, CharlesThesis}.  }}
{\includegraphics[width=78mm]{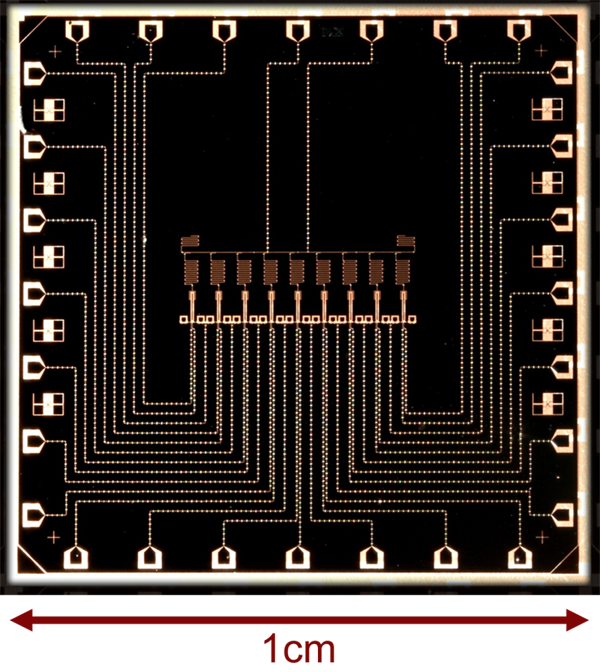}}
\end{figure}

\newpage

\section{\textcolor{SECTIONCOLOR}{2. Spectroscopy based on fundamental postulate of quantum mechanics}}

\vspace{-3pt}
According to the time-independent Schr$\ddot{\text{o}}$dinger equation, the time evolution of the state of the system $|\psi(t)\rangle$  is given

\begin{equation}
|\psi(t)\rangle = \sum_{\alpha}C_{\alpha}e^{-iE_{\alpha}t}|\phi_{\alpha}\rangle,
\end{equation}

\noindent
where $C_{\alpha}=\langle \phi_{\alpha}|\psi_0\rangle$, and $|\psi_0\rangle$ is the initial state. On the other hand,  an observable in the energy basis can be written as

\begin{equation}
\hat{O}=\sum_{\alpha,\alpha'}O_{\alpha',\alpha}|\phi_{\alpha'}\rangle \langle \phi_{\alpha}|,
\end{equation}

\noindent
where $O_{\alpha',\alpha}= \langle \phi_{\alpha'}|\hat{O}|\phi_{\alpha}\rangle$ and accordingly its expectation value is:

\begin{equation}
O(t)=\langle \psi(t)|\hat{O} |\psi(t) \rangle=\sum_{\alpha,\alpha'}O_{\alpha',\alpha}C_{\alpha}C_{\alpha'}^*e^{-i(E_\alpha-E_{\alpha'})t}.
\end{equation}

\noindent
For our spectroscopy purpose, we are interested in energies $E_\alpha$ and not energy differences $E_\alpha-E_{\alpha'}$. The above relation suggests that when measuring any observable one will generally end up with energy differences. It is not obvious \textit{a priori} how to avoid energy differences. However, proper choices of observables and initial states can help to overcome this issue, enabling extraction of eigen-energies. A key observation here is that one should somehow fix $E_{\alpha'}$ to a specific energy, i.e., \textit{a reference energy}. See the schematic in Fig.\,S2.

\vspace{-10pt}

\subsection{\textcolor{SUBSECTIONCOLOR}{2.1 Energy differences vs. absolute energies}}

\vspace{-3pt}

We illustrate the method by considering first a simple example of two coupled harmonic oscillators described by the tight-binding Hamiltonian
\begin{equation}
H_{\text{dimer}}=\omega (a^\dagger_1a_1+a^\dagger_2a_2)-J (a^\dagger_1a_2+H.c.),
\end{equation}

\begin{figure}[h]
\includegraphics[width=180mm]{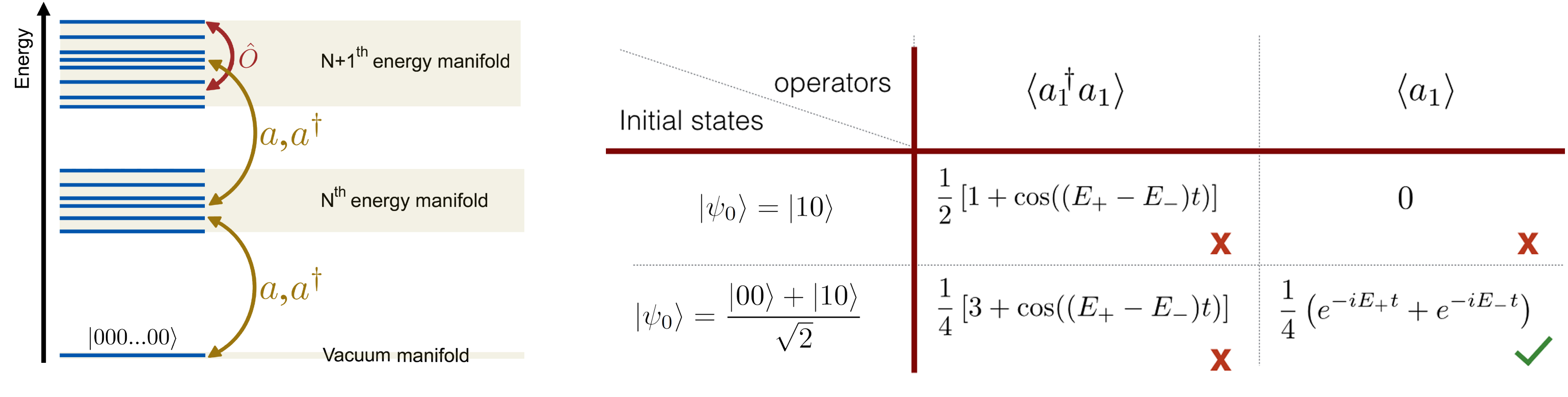}
\caption{ \textbf{(a)} The schematic shows that generally operators, such as $\hat{O}$ connect different levels, and hence in their observables one would see the energy differences of the levels. The raising $a^\dagger$ and lowering $a$ operators connect states in different manifolds. It is a fortunate coincidence that there is a manifold with only one state in it, the vacuum manifold, which can act as a energy reference. The initial states that are superposition of vacuum state and some other state are necessary for having a functional protocal. \textbf{(b)} We show expectation values of different operators for two different initial states, associated with the example that we used. Note that both proper initial states and choice of operators are needed to have a useful protocol.}
\label{fig:table}
\end{figure}

\vspace{8pt}
\noindent
where $\omega$ is the frequency of the oscillators, $J$ is the hopping rate, $a_1$ and $a_2$ are bosonic annihilation operators acting on the first and the second oscillators, respectively. The single-photon eigenstates are $|\phi_\pm\rangle = (|10\rangle\pm|01\rangle)/\sqrt{2}$ with the eigen-energies  $E_\pm=\omega \pm J$, where $|10\rangle = a^\dagger_1|00\rangle$, $|01\rangle=a^\dagger_2|00\rangle$ and $|00\rangle$ is the vacuum. Different choices of initial states and observables are shown in Fig.\,\ref{fig:table} (right panel). One quickly realizes that $a^\dagger a$ would have energy differences and not suitable; but $a$ might, if the proper initial state is chosen. A proper initial state would be the superposition of the relevant number state (here the $|10\rangle $ state) that has one particle and the vacuum $|00\rangle $. The vacuum here serves as the appropriate reference state and has energy $E_{\alpha'}=0$. Since $a$ is non-Hermitian an hence not an observable, $a$ cannot be measured directly. However, it can be easily inferred from its Hermitian "quadratures" $\sigma^X$ and $\sigma^Y$  as $\langle a\rangle=\langle\sigma^X\rangle+i\langle\sigma^Y\rangle$, where $\sigma^X=|1\rangle\langle 0|+|0\rangle\langle 1|$ and $\sigma^Y=i|1\rangle\langle 0|-i|0\rangle\langle 1|$ are Hermitian and hence observable.

\newpage
\subsection{\textcolor{SUBSECTIONCOLOR}{2.2 Single-particle spectroscopy: generalization}}

Now let us consider a general particle-conserving interacting Hamiltonian $H$ in a lattice and assume first that there is only one particle in the system. As before, we choose an initial state as a product state of the form

\begin{equation}
|\psi_0\rangle_n = |0\rangle_1 |0\rangle_2 ...\left(\frac{|0\rangle_n+|1\rangle_n}{\sqrt{2}}\right) ...|0\rangle_{N-1}|0\rangle_{N}  =\frac{1}{\sqrt{2}}\left(|\text{Vac}\rangle +| \mathbf{1}_n\rangle\right),
\end{equation}

\vspace{8pt}
\noindent
where $N$ is the number of sites, $n\in\{1,2,..,N\}$, $|\text{Vac}\rangle \equiv |0\rangle_1 |0\rangle_2 ...|0\rangle_N $ is the vacuum state and $|\mathbf{1}_n\rangle = |0\rangle_1 |0\rangle_2 ... |\boldmath{1}\rangle_n ...|0\rangle_N$ is a one-photon Fock state. The state at time $t$ is

\begin{equation}
|\psi (t)\rangle_n = \frac{1}{\sqrt{2}}\left(|\text{Vac}\rangle +\sum_{\alpha}C_{\alpha,n}e^{-iE_{\alpha}^{(1)}t}|\phi^{(1)}_{\alpha}\rangle\right),
\end{equation}

\vspace{8pt}
\noindent
where $\alpha\in\{1,2,..,N\}$, $C_{\alpha,n}=\langle \phi^{(1)}_{\alpha}|\mathbf{1}_n\rangle$ and $|\phi^{(1)}_{\alpha}\rangle$ is the one-photon eigenstate with the eigen-energy $E^{(1)}_{\alpha}$. The expectation value of $\chi_1(n) \equiv \langle\sigma^X_n\rangle+i\langle\sigma^Y_n\rangle$ takes the form

\begin{equation}
\chi_1(n) = \frac{1}{2}\sum_{\alpha}|C_{\alpha,n}|^2e^{-iE_{\alpha}^{(1)}t}.
\end{equation}

Non-vanishing peak amplitudes $|C_{\alpha,n}|^2>0$ can be assured by varying the initial state, i.e. varying $n$, to span the space of the single-photon manifold.

\subsection{\textcolor{SUBSECTIONCOLOR}{2.3 Two-particle spectroscopy: generalization}}

The initial states are product states of the form

\begin{equation}
|\psi_0\rangle_{n,m} = |0\rangle_1 |0\rangle_2 ...\left(\frac{|0\rangle_n+|1\rangle_n}{\sqrt{2}}\right) ...\left(\frac{|0\rangle_m+|1\rangle_m}{\sqrt{2}}\right) ...|0\rangle_{N-1}|0\rangle_N  =\frac{1}{2}\left(|\text{Vac}\rangle +|\mathbf{1}_n, \mathbf{1}_m\rangle \right)+\frac{1}{2}\left(| \mathbf{1}_n\rangle+|\mathbf{1}_m\rangle\right),
\label{eq:two_photons_init}
\end{equation}

\vspace{8pt}
\noindent
where $n\neq m$, and  $m,n\in\{1,2,..,N\}$ and $|\mathbf{1}_n,\mathbf{1}_m\rangle = |0\rangle_1 |0\rangle_2 ...|1\rangle_n ...|1\rangle_m ...|0\rangle_{N}$ are the two-photon Fock states. The state at time $t$ is

\begin{equation}
|\psi (t)\rangle_{n,m} = \frac{1}{2}\left(|\text{Vac}\rangle +\sum_{\beta}C_{\beta,(n,m)}e^{-iE_{\beta}^{(2)}t}|\phi^{(2)}_{\beta}\rangle\right)+
\frac{1}{2}\left(\sum_{\alpha}(C_{\alpha,n}+C_{\alpha,m})e^{-iE_{\alpha}^{(1)}t}|\phi^{(1)}_{\alpha}\rangle\right),
\end{equation}

\noindent
where $\beta\in\{1,2,.., \frac{1}{2}N(N+1)\}$, $|\phi^{(2)}_{\beta}\rangle$ is an energy eigenstate in the two-photon manifold with the corresponding energy $E^{(2)}_{\beta}$ and $C_{\beta,(n,m)}=\langle \phi^{(2)}_{\beta}|\mathbf{1}_n,\mathbf{1}_m\rangle$. A generalized two-photon lowering operator can be constructed as

\begin{equation}
\chi_2(n,m) \equiv \langle\sigma^X_n\sigma^X_m\rangle - \langle\sigma^Y_n\sigma^Y_m\rangle  + i \langle\sigma^X_n\sigma^Y_m\rangle  + i \langle\sigma^Y_n\sigma^X_m\rangle .
\end{equation}

\vspace{8pt}
\noindent
This operator measures the phase difference between the vacuum and the two-photon state, while projecting out the one-photon component to avoid measuring the energy differences $E^{(2)}_{\beta}-E^{(1)}_{\alpha}$. Its expectation value takes the form

\begin{equation}
\chi_2(n,m)= \frac{1}{4}\sum_{\beta}|C_{\beta,(n,m)}|^2e^{-iE_{\beta}^{(2)}t}.
\end{equation}

One might observe that with our choice of initial states, we do not directly cover all the space in the two-photon subspace since we did not include double-occupancy states such as $|\mathbf{2}_n\rangle \equiv |0\rangle_1 |0\rangle_2 ...|2\rangle_n ...|0\rangle_N$.  However, in the soft-core limit $U/J=3.5$ where we operate, all 45 two-photon eigenstates $|\phi_{\beta}^{(2)}\rangle$ have appreciable overlap with our choice of initial states. Therefore, their energies can be measured as shown in the main text. As one get to the hardcore limit with $U/J\to \infty$, mainly 36 out of 45 eigen-energies will be picked up by $\chi_2(n,m)$, which again are all the physically relevant ones to probe the physics of the system in this regime.
\newpage

\begin{figure}[h]
\includegraphics[width=150mm]{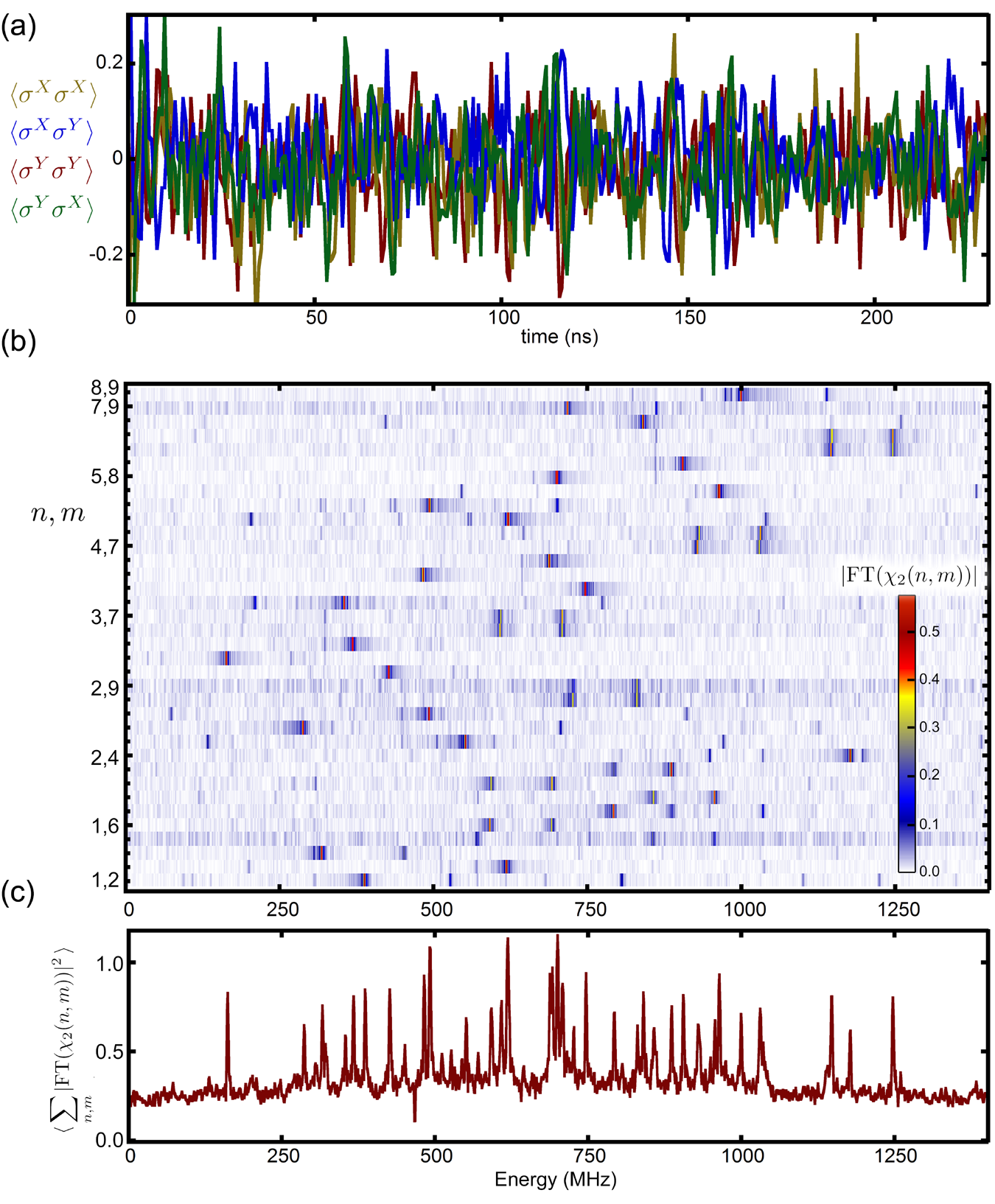}
\caption{\textbf{Spectroscopy of energy levels in the two-photon manifold} The protocol for taking data in the two-photon manifold is very similar to the method illustrated in Fig.\,(1) of the main text. \textbf{(a)} A typical time-domain measurement of the two-point correlations that are needed for constructing $\chi_2(n,m)$. For this data set $n=5$ and $m=7$. Similar measurements are done for every $m,n\in\{1,2,..,N\}$ with $n\neq m$. \textbf{(b)} The magnitude of the Fourier transform of $\chi_2(n,m)$ for all $m$ and $n$ choices with $m,n\in\{1,2,..,N\}$ and $n\neq m$. \textbf{(c)} Average of Fourier transforms presented in \textbf{(b)}. }
\end{figure}

\vspace{65pt}
\newpage

\subsection{\textcolor{SUBSECTIONCOLOR}{2.4 Computation of the Participation Ratio}}

Here, we discuss in details how the participation ratios in the two-photon manifold are measured and computed. In the Bose-Hubbard Hamiltonian

\vspace{9pt}
\begin{equation}
H_{BH} = \sum\limits_{n=1}^{9} \mu_n a^{\dagger}_{n}a_{n}+ \frac{U}{2}\sum\limits_{n=1}^{9} \,a^{\dagger}_{n}a_n(a^{\dagger}_{n}a_n-1)+J\sum_{n=1}^8 a^{\dagger}_{n+1}a_n+a^{\dagger}_{n}a_{n+1},
\end{equation}

\vspace{9pt}
\noindent
we set $J/2\pi=50$\,MHz. By design of the chip, $U$ is fixed $U/2\pi=175$\,MHz. We realize a quasi-periodic potential by setting $\mu_n=\Delta \cos(2\pi n b)$, where $b=(\sqrt{5}-1)/2$. We vary $\Delta$ from 0 to $\Delta/2\pi=300$\,MHz. Choosing the inverse of the so-called golden ratio for $b$ stems from the fact that this irrational number is considered to be "very" irrational number, meaning that approximating it in terms of ratio of integers involves large numbers. Nevertheless, in setting the parameters in the lab, there is no meaningful distinction between rational and irrational numbers. We chose this number just to reduce the chance of commensurability with the lattice, which could have been the case if we chose a number close to 0.5 or 0.33.

Initial states are made by placing two qubits ($Qn$ and $Qm$) in the superposition of the $|0\rangle$ and $|1\rangle$ state. We measure two-point correlations and construct $\mathcal{\chi}_2(n,m)\equiv \langle\sigma^X_n\sigma^X_m\rangle -\langle\sigma^Y_n\sigma^Y_m\rangle+i\langle\sigma^X_n\sigma^Y_m\rangle+i\langle\sigma^Y_n\sigma^X_m\rangle$. We consider all pairs of qubits for the initial states $n,m \in \{1,2,...,9\}$. In the two-photon energy manifold, there are $36$ single \,(e.g. $|\mathbf{1}_3,\mathbf{1}_7\rangle=|001000100\rangle$) and 9 double occupancy states\,(e.g. $|\mathbf{2}_5\rangle=|000020000\rangle$), which gives ${9 \choose 2}+{9 \choose 1}=45$ energy levels. In the two-photon energy manifold of Egn.\,(2), we create all 36 single-occupation initial states:

\vspace{9pt}
\begin{equation}
|\psi_0\rangle_{n,m} \equiv |0\rangle_1 |0\rangle_2 ...\left(\frac{|0\rangle_n+|1\rangle_n}{\sqrt{2}}\right) ...\left(\frac{|0\rangle_m+|1\rangle_m}{\sqrt{2}}\right) ...|0\rangle_{N-1}|0\rangle_N
\end{equation}

\vspace{9pt}
\noindent
In the average of the magnitude of FT of data (summed over all these 36 initial states), we identify 45 peaks. This constitute $\{E_\alpha\}$, a set of 45 eigen-energies of the Hamiltonian. Note that the number of initial states that one can begin with is technically infinitely many. Using more initial states only adds to the confidence for the detected peaks. Since all the eigen-energies of the Hamiltonian has been identified, by choosing one of the initial states, \,(e.g. $|001000100\rangle$), we can see how extended it is in the energy landscape. This is done by considering the FT of the data and reading the magnitude of FT for all $E_\alpha$ values. Therefore in the expansion

\vspace{6pt}
\begin{equation}
|\psi_0\rangle_{n,m}=\sum_{\alpha}{C_{\alpha,(n,m)}|\phi_{\alpha}\rangle}
\end{equation}
\vspace{9pt}

\noindent
we now know $\{P_{\alpha,(n,m)}\}=\{|C_{\alpha,(n,m)}|^2\}$. Next, we want to extract the expansion of Fock states $|1_n\rangle$ from the expansions of our initial states that involved two photons. This is simply done by adding probabilities of the initial states which share one of the excited qubits  $P_{\alpha,n}=\sum_m P_{\alpha,(n,m)}$. Now we can compute

\vspace{6pt}
\begin{equation}
PR_{\text{Energy}}(n) \equiv \frac{1}{\sum\limits_{\alpha}P_{\alpha,n}^2},
\end{equation}
\vspace{6pt}

\noindent
which is a measure of the spread of a real-space localized state in energy landscape. Note that since $\sum\limits_{\alpha} P_{\alpha,n}=1$,  the $PR_{\text{Energy}}$ is the simplest non-trivial moment of the probability distribution, which tells us about the spread of wavefunctions. The expansion of energy eigenstate in real space $|\phi_{\alpha}\rangle=\sum_{n}{C_{\alpha,n}|\mathbf{1}_n\rangle}$ is readily done by summing over $n$
\vspace{6pt}
\begin{equation}
PR_{\text{space}}(\alpha) \equiv \frac{1}{\sum\limits_{n}P_{\alpha,n}^2}.
\end{equation}
\vspace{6pt}

\begin{figure}[h]
\floatbox[{\capbeside\thisfloatsetup{capbesideposition={right,top},capbesidewidth=3.7cm}}]{figure}[\FBwidth]
{\caption{\textbf{Computation of the Participation Ratio in the single-photon manifold.} \textbf{(a)} This is the same plot as Fig.1\,(c) of the main text and for the ease of comparison is represented in here. \textbf{(b)} After identification of the peaks, we ask what is the magnitude of FT\,(false color) at each site\,(vertical axis) for a given eigen-energy\,(horizontal axis). We normalize the amplitudes such that $\sum_n P_{\alpha,n}=1$ and $\sum_\alpha P_{\alpha,n}=1$.  }}
{\includegraphics[width=130mm]{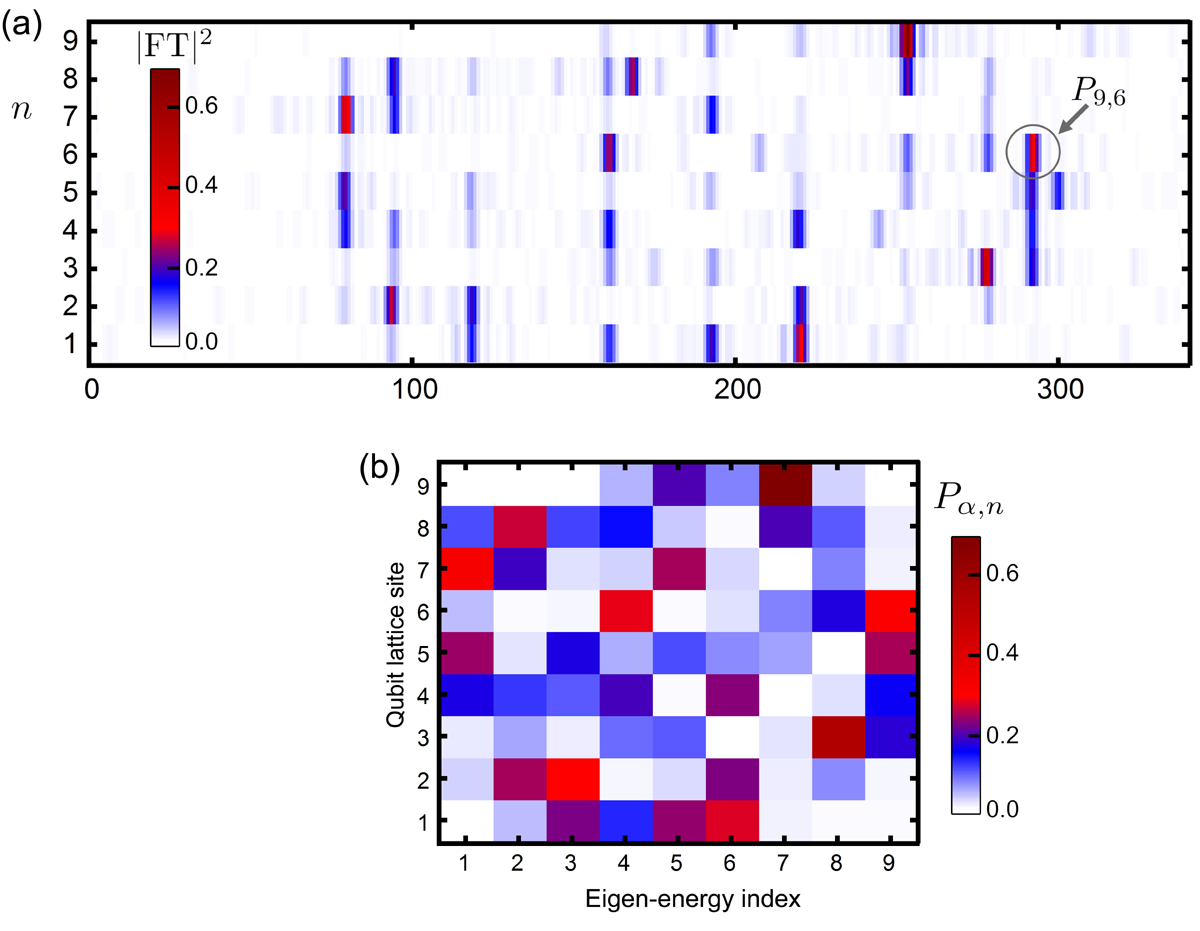}}
\end{figure}

\subsection{\textcolor{SUBSECTIONCOLOR}{2.5 Resolving the spectrum for large Hilbert-spaces}}

In this section, we address the main technical challenge that arises when dealing with a large many-body system. What sets the limit on how small and large energy differences we can resolve?

\vspace{8pt}

In the current work the main limitation comes from the data collecting rate. Each panel in the two-photon manifold (Fig 3,4, and 5), where we change $\Delta$ in about 20 steps, took about 2 days to be collected. Recall that we generated all single-occupancy two-photon states for every realization of the Hamiltonian. This is not a fundamental limit and one can take data for longer time, but we impose this limit ourselves. Instead of taking data with this rate for a longer time, we are addressing the core of this issue and working on improving our data taking rate. There are indications that with fast resetting techniques and better data streaming, we can improve the data taking rate by two orders of magnitude. These ideas will be tested and implemented in the future generations of our devices. These methods would help to push the limitations for implementation of our spectroscopy method to be limited only by the coherent time of the system.

When the system size $N$ is increased, missing some energy peaks in the measurement will eventually become unavoidable because the resolution in the energy Fourier spectrum is fixed by the coherence time of the system. Here, we analyze what happen to level statistics when some levels are missing. To study the deviation from the ideal level spacing distributions, we use the Kullback-Leibler divergence and estimate the efficiency of our method

\begin{equation}
      \label{eq:KLDiv}
            D_{KL}(P\| Q)=\sum_l P(r_l)\log\left(\frac{P(r_l)}{Q(r_l)}\right).
\end{equation}

\vspace{8pt}
\noindent

The KL divergence is close to zero when the two distributions $P$ and $Q$ are close. In Fig.\,S5(a) we show the number of missing levels for different parameter regimes and a fixed resolution of $1MHz$. Even for a finite size $N=18$, it is interesting that one misses more levels close to the critical point of the AA model $\Delta\approx 2J$. This is expected because close to the critical point, some levels cluster and the statistics is neither Poisson nor GOE. Therefore, for a finite resolution of the Fourier spectrum, there would be more missing levels. Fig.\,S5(b) shows a second step to test the efficiency of our method. We calculate numerically the KL divergence for a fixed resolution which leads to different number of missing levels for different parameters.

\begin{figure}[h]
\floatbox[{\capbeside\thisfloatsetup{capbesideposition={right,top},capbesidewidth=5cm}}]{figure}[\FBwidth]
{\caption{\textbf{The effect of missing levels in level statistics} We numerically simulate a system of two interacting particles in a lattice with $N=18$ sites and $J/2\pi=50MHz$. \textbf{(a)} For a fixed resolution of $1MHz$, we calculate the percentage of missing levels for different parameter regimes. \textbf{(b)} To determine the effect of missing levels on the statistics we use the Kullback-Leibler (KL) divergence between the measured distribution (which had missing levels) to the distribution without missing levels.}}
{\includegraphics[width=115mm]{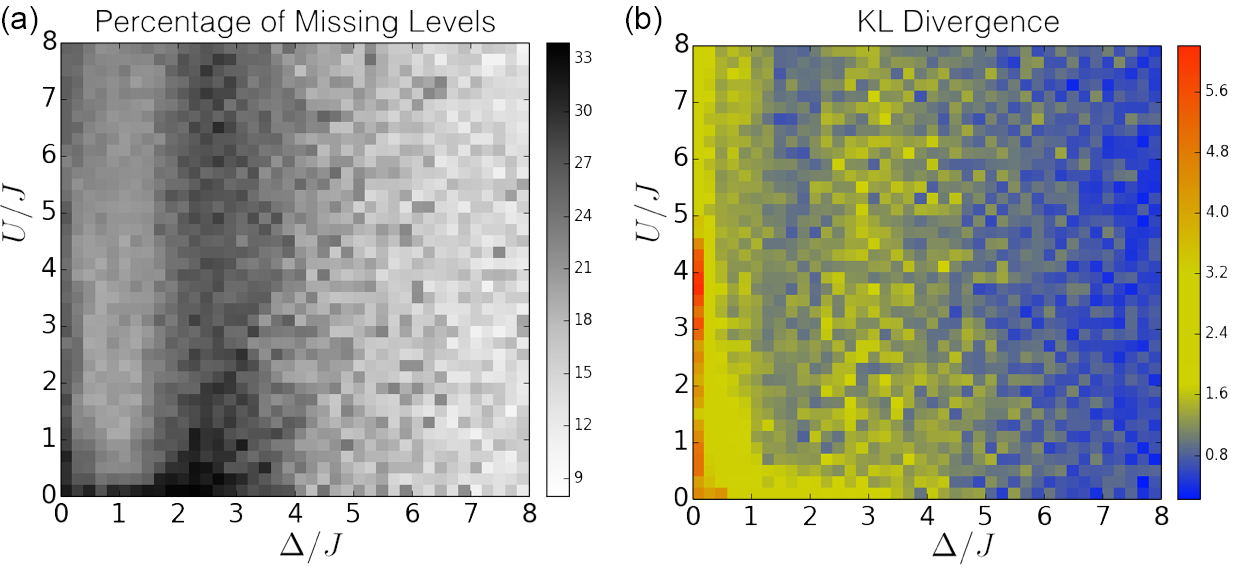}}
\end{figure}

%\newpage
\section{\textcolor{SECTIONCOLOR}{3. Two-point quantum correlations}}

Here we discuss the details of the two-point correlation measurements and also provide the numerical simulations. In the Bose-Hubbard Hamiltonian

\vspace{9pt}
\begin{equation}
H_{BH} = \sum\limits_{n=1}^{9} \mu_n a^{\dagger}_{n}a_{n}+ \frac{U}{2}\sum\limits_{n=1}^{9} \,a^{\dagger}_{n}a_n(a^{\dagger}_{n}a_n-1)+J\sum_{n=1}^8 a^{\dagger}_{n+1}a_n+a^{\dagger}_{n}a_{n+1},
\end{equation}

\vspace{9pt}
\noindent
we set $J/2\pi=50$\,MHz. By design of the chip, $U$ is fixed $U/2\pi=175$\,MHz. We realize a quasi-periodic potential by setting $\mu_n=\Delta \cos(2\pi n b)$, where $b=(\sqrt{5}-1)/2$, and vary $\Delta$.

\vspace{9pt}
In the data presented in Fig.\,(5) of the main text, The initial states are made by placing two qubits ($Qn$ and $Qm$) in the superposition of the $|0\rangle$ and $|1\rangle$ states. We measure $\mathcal{S}_{m,n}=|\langle\sigma_m^1 \sigma_n^2\rangle-\langle\sigma_m^1\rangle \langle\sigma_n^2\rangle|$, where $\sigma^1,\sigma^2 \in \{\sigma^X, \sigma^Y\}$ and $m,n \in \{1,2,...,9\}$, for all $m$ and $n$ combinations and choices of Pauli operators. The number of pairs of qubits that one can pick for exciting initially and measuring the two-point correlation is ${9 \choose 2}=36$. The total number of choices for Pauli operators is 4 ($XX,YY,XY,YX$), which means that $36\times4=144$ distinct $\mathcal{S}_{m,n}$ are measured. Next, we average $\mathcal{S}_{m,n}$ over time, from 0 to 250 ns, and over the 4 choices of the Pauli operators, and over all qubit pair combinations with the same $|m-n|$. This gives ${\mathcal{\tilde{S}}_{m,n}}$ for a given realization of Hamiltonian (a given $\Delta/J$) as a function of $|m-n|$.

\vspace{20pt}

\begin{figure}[h]
\floatbox[{\capbeside\thisfloatsetup{capbesideposition={right,top},capbesidewidth=3.5cm}}]{figure}[\FBwidth]
{\caption{\textbf{Numerical calculation of the quantum correlations} We consider $J/2\pi=50MHz$, which fixes the interaction $U/J=3.5$. The time-averaged(from 0 to 250 ns) correlation $\mathcal{S}_{m,n}=|\langle\sigma_m^1 \sigma_n^2\rangle-\langle\sigma_m^1\rangle \langle\sigma_n^2\rangle|$ for $b=(\sqrt{5}-1)/2$ and a system size $N=9$ is \textbf{(a)} experimentally measured (same as Fig.\,5 in the manuscript), and \textbf{(b)} numerically computed. }}
{\includegraphics[width=140mm]{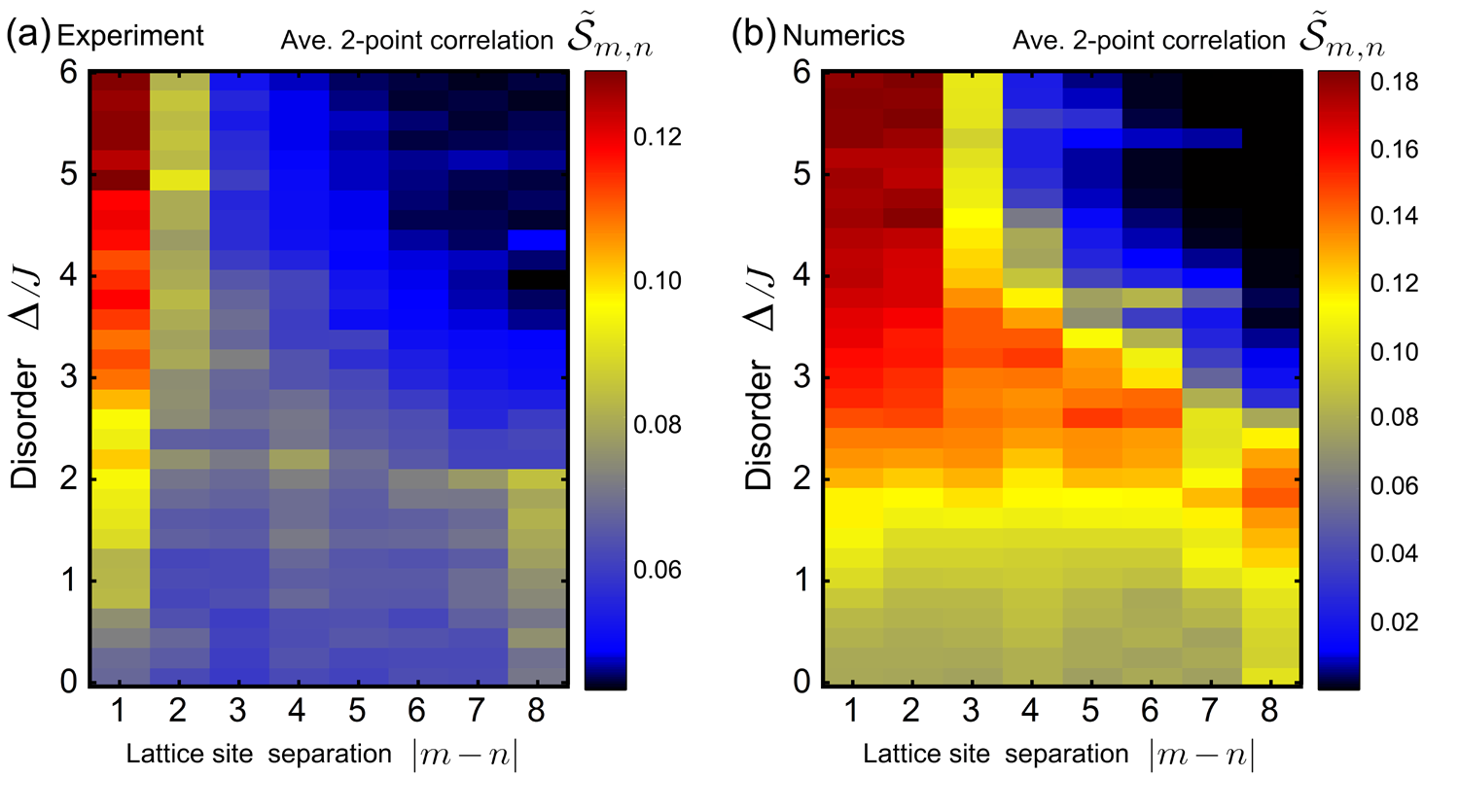}}
\end{figure}

\newpage

\section{\textcolor{SECTIONCOLOR}{4. Hamiltonians used in this work }}

\subsection{\textcolor{SUBSECTIONCOLOR}{4.1 Mapping the 2D quantum Hall model to the 1D Harper model}}

\begin{figure}[h]
\includegraphics[width=180mm]{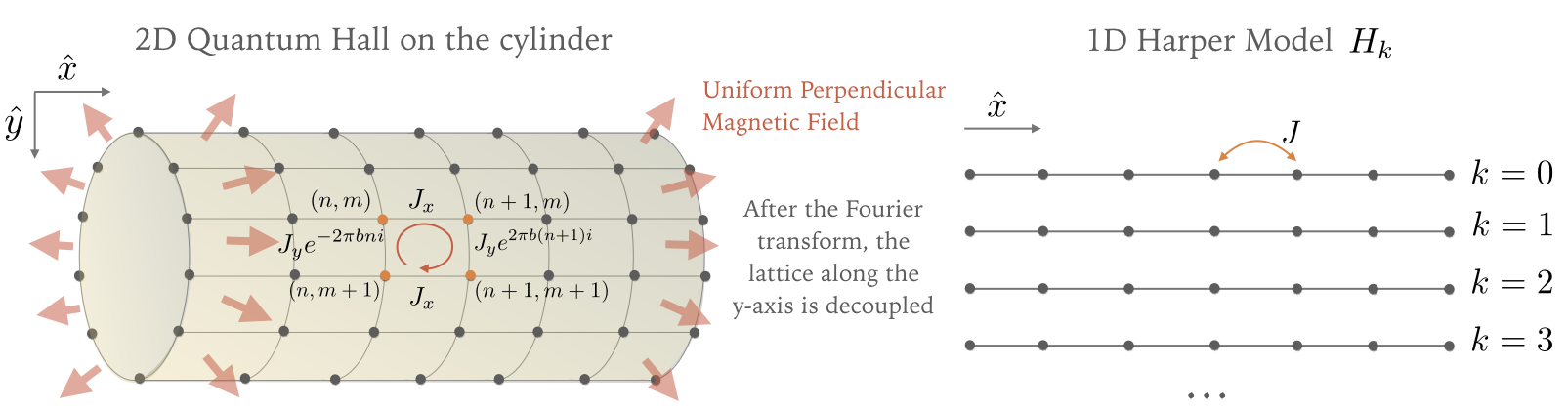}
\caption{Illustration of the 2D Quantum Hall model and its mapping to the 1D Harper model. }
\label{fig:2D1D}
\end{figure}

An electron moving in a 2D lattice with a perpendicular magnetic field $b$ is described by the quantum Hall model,

\begin{equation}
H_{\text{IQH}} =  J_x\sum_{n,m}\left(\hat{a}^\dagger_{n,m} \hat{a}_{n+1,m} + H.c.\right)+J_y\sum_{n,m}\left(\hat{a}^\dagger_{n,m}\hat{a}_{n,m+1}e^{2\pi b n} + H.c. \right)
\label{eq:iqh}
\end{equation}

where $J_x$ and $J_y$ are hopping strength along x and y axes respectively, see Fig.\ref{fig:2D1D}. For a periodic boundary condition on the y-direction, one can define the quantum fourier transform $\hat{a}^\dagger_{n,m}=\sum_{k}e^{-ikm}\hat{a}^\dagger_{n,k}$. Substituting this to Eq.\ref{eq:iqh}, we get $H_{IQH}=\sum_{k}H_k$, where

\begin{equation}
H_k = 2J_y \sum_{n}\cos(2\pi b n + k)\hat{a}^\dagger_{n,k}\hat{a}_{n,k}+J_x\sum_{n}\left(\hat{a}^\dagger_{n,k}\hat{a}_{n+1,k}+H.c. \right)
\end{equation}

The 1D harper model is then achieved by dropping the index $k$ in $\hat{a}_{n,k}$ and replacing $2J_y$ and $J_x$ with $\Delta$ and $J$, respectively. We set $k=0$ in the main text. Edge states in the Harper model has been studied in\,\cite{Kraus2012}.

\subsection{\textcolor{SUBSECTIONCOLOR}{4.2 Aubry-Andre model}}

In the absence of interactions (or at single-particle level), the Eqn.\,3 of the main text is the celebrated Aubry-Andre (AA) model\cite{Huse2013, Altshuler2014}. We compare this model with the well-studied Anderson model. In 1D or 2D Anderson model, any amount of disorder would localize the entire system and there is no phase transition, or transition is at zero disorder. However, in the 3D Anderson model there is a localization-delocalization transition, where a mobility edge appears\,\cite{Abrahams1979}. Similar to the 3D Anderson model, the 1D AA model exhibits a localization-delocalization transition. In the AA model,  when $\Delta=2J$,  all the eigenstates are fully localized (delocalized) for $\Delta>2J$ ($\Delta<2J$) and the localization length is independent of the energy and is solely determined by the ratio between $\Delta$ and $J$\,\cite{Altshuler2014}. Therefore, the AA model does not exhibit mobility edge \cite{DasSarma2015}. The main difference between the 3D Anderson model and the AA model is that in the AA model, the delocalized phase is characterized by ballistic transport, i.e., scattering events are rare\cite{Huse2013}. This implies a ballistic spreading of an initial state localized at a given site.

Now let us discuss briefly the effect of interactions in the AA model (Eqn.\,2 of the main text) and a discussion about the signatures of the mobility edge. In the interacting case, one can use the basis of single-particles states of the AA model and in this representation, the interaction act as a hopping term in energy space, allowing transitions between different single-particle states\,\cite{Altshuler2014}. In systems with a single-particle mobility edge, like 3D Anderson model, the interaction couples localized and delocalized states\cite{DasSarma2015}. This happens because for a given  strength of disorder, localized and delocalized states coexist. In the case of the AA model, as there is not mobility edge in the noninteracting case, the interaction just couples single-particle states which are either localized or delocalized. The signatures of the mobility edge observed in the experiment can be explained as a consequence of the interaction between the two particles in the lattice. When they are located at a distance smaller that the single-particle localization length, they form a bound state that spreads ballistically, as it has been reported in the literature \cite{Khomeriki2012}. When the particles are far apart, i.e., at a distance longer than the localization length, they remain localized. This argument explain the coexistence of localized and delocalized states in the interacting model for a fixed disorder and provides an explanation for the signatures of the mobility edge seen in Fig.\,(3).